\documentstyle [12pt]{article}
\title{Electronic structure and properties of pure and doped $\epsilon$-FeSi
from ab-initio local density theory.}
\author{ 
T. Jarlborg\\ 
DPMC, University of Geneva, \\
24 Quai Ernest Ansermet, CH-1211 Geneva 4, Switzerland \\}

\begin{document}
\baselineskip 5 mm
%\begin{titlepage}
\maketitle   
\begin{abstract}
Local density calculations of the electronic structure
of $FeSi, FeSi_{1-x}Al_x$ and $Fe_{1-x}Ir_xSi$ systems in the B20 structure
are presented.
Pure FeSi has a semi-conducting gap of 6 mRy at 0 K.
 Effects of temperature (T) in terms of electronic and
vibrational excitations are included. Various measurable properties,
such as magnetic susceptibility $\chi(T)$, electronic specific heat $C(T)$, 
thermoelectric power $S(T)$, relative variations in resistivity $\rho(T)$, 
and peak positions in the density-of-states (DOS)
are calculated. The feedback from vibrational disorder onto the electronic
structure is found to be essential for a good description of most properties, although
the results for $S(T)$ in undoped FeSi can be described up to about 150 K without considerations
of disorder. Above this T, only the filling of the gap due to disorder accompanied 
by exchange enhancement, can explain the large susceptiblity. The overall
good agreement with experimental data for most properties in doped and pure FeSi
suggests that this system is well described by LDA even at large T. Doped FeSi can be described quite
well from rigid-band shifts of the Fermi energy on the DOS of pure FeSi. Spin-polarization in
Ir doped FeSi leads to a semi-metallic magnetic state at low T.

(Submitted to Phys. Rev. B)
\end{abstract}

PACS:  71.20-b, 71.30+h.

%\end{titlepage}
\section{Introduction}
\vspace{0.5cm}

The FeSi-system has been much studied since the early work of Jaccarino
{\it et al.} \cite{jac} which showed that the magnetic susceptibility and
specific heat have unusually strong T-dependence. A renewed interest
in this material is reflected in the large number of experimental
and
theoretical works that have been published in the last few years \cite{mat}-\cite{hun}.
Part of the interest is technological. A large thermopower in combination with low resistivity
and thermal conductivity of a semi-conductor, could be exploited in form of
a Peltier refrigerator. Experimental studies indicate that FeSi has optimal properties for this 
in the temperature range near 100 K, although they are not large enough for practical applications 
\cite{sal}. But most studies deal with the physics of FeSi, to find the reasons for 
its unusual properties.
Much of the discussion concerns the properties at high T, where many properties 
cannot be explained by traditional means. At low T,
resistivity, $\rho(T)$, and optical measurements agree with 
band theory in that FeSi is a very narrow gap semi-conductor, whereas
the opinions diverge on what happens at large T (from $\sim$ 150 K and above). Most properties
indicate that FeSi is metallic and exchange enhanced at large T,
which seem incompatible with
the gap being 4-7 mRy as obtained from the band results based on the local
density approximation (LDA) $\cite{mat}-\cite{ani}$.
Some recent works propose that FeSi is exotic in the sense that it
cannot be described by LDA at large T, because of strong correlation,
Kondo screening or Mott localization \cite{chr,don,ani}.
 
Optical data agree well with
LDA bands and show a gap of about 5 mRy at T=0, but this gap disappears 
already at about 200 K \cite{sch}. This work claimed also that there
is a 'missing intensity' so that the total intensity in the filled 
gap at large T, was not compensated by the intensity from above gap
in the low-T data. This conclusion was later refuted in new measurements
by Degiorgi {\it et al.} \cite{deg}, although the unusual disappearence of
the gap made these authors propose a Mott-localization. The optical
data cannot be explained in simple terms of normal transition probabilities
using the LDA band structure, and this inspired Fu {\it et al.} \cite{don} to propose 
that strong correlation
in the form of  Kondo-like excitations causes disorder of the band structure.
The large value of $\chi$ and $C$ at large T can be modelled by very large
and narrow DOS-peaks on both sides of the gap \cite{jac,man}, but
such peaks need to be much larger than, and are incompatible with, what is found
from band theory \cite{mat}.

On the other hand, many properties at low T as well as in doped
FeSi agree well with the band results. In addition to the resistivity and 
optical data as mentioned earlier,  the measured thermopower 
\cite{sal} is very well reproduced
by Boltzmann theory using LDA bands for slight hole doping \cite{jar}.
Moreover,
the pressure dependence of the magnetic susceptibility has been found to
follow the calculated variation of the band gap $\cite{gre}$.
Recent studies of doped FeSi conclude likewise that low-T properties
are well described by LDA results, while at larger T the systems look more like
what is typically seen in
Kondo insulators with heavy mass features \cite{ditusa}. 

A different picture of the FeSi system comes from LDA calculations
which consider thermal disorder \cite{jar2,jar3}. The effective DOS at $E_F$ increases more
rapidly with T than what is obtained from a Fermi-Dirac smearing only.
The calculated $\chi(T)$ is in
satisfactory agreement with experiment in this case \cite{jar3}, and
the band filling is simply a result of the thermal disorder.
The DOS peaks near the gap become broadened and fill gradually the gap, much like what has been
observed  in angular resolved
photoemission \cite{par} and optical measurements \cite{sch,deg}. Qualitatively, 
the effect of disorder on many DOS-dependent properties, is similar
to the effect of Fermi-Dirac smearing. But the temperature scale has to be redefined.
This is exemplified by the T-variation of the quadrupole splitting 
on Fe, when it is interpreted via a Fermi-Dirac smearing of the DOS determined for 
the ordered lattice; a good fit to experiment was found when the temperature scale was 
renormalized by a factor of 10 \cite{chr}.

At very low T it is likely that impurities play a role in determining the properties of pure FeSi,
for instance in the often seen upturn in $\chi$ at the lowest T.
Measurements of $\chi, C$ and $\rho$ below about 10 K, show anomalies which can be
interpreted in terms of Anderson-localization of impurity states \cite{hun}.

The purpose of this paper is to investigate how the 
electronic structure and various
properties are affected by the thermal disorder and doping. 
 Part of these results have been published
in shortened form \cite{jar,jar2,jar3}, while here more complete results
are presented. Further, new results of Al- and Ir-doped FeSi are presented,
as well as a more detailed description of the methods of calculation.

\section{Methods of calculation.}

\subsection{Electronic structure.}

The cubic compound $\epsilon$-FeSi has eight atoms per unit cell and its
crystal structure is B20, which can be viewed as a distorted rocksalt structure \cite{mat}.
The structure is fairly close packed with a "touching sphere" 
volume ratio of 62 percent compared to 64 percent for bcc.
For the band calculations we use the self-consistent, semi-relativistic,
Linear Muffin-Tin Orbital (LMTO) method $\cite{lmt}$ with LDA potentials \cite{lda}. 
The basis set included up to f-states for
the 8 atom unit cell. Supercell calculations with 64 atoms/cell (where the basic cell is doubled 
in the x-, y- and z-direction) included s-,p- and d-states in the direct basis,
while the f-states were included only in the 3-center terms. 
The calculations contain the so-called "combined correction" terms to the overlaping spheres, 
but no other non-spherical potential corrections are included.

The use of standard LMTO in this work is motivated by the fact that none of the studied properties
require a "full-potential" method.  Calculations of elastic constants
in bcc and fcc metals including full-potential corrections, based on calculations of 
differences in total energies for different distortions, showed that these corrections
were essential for the value of the elastic constants, but the effect on the DOS is hardly
visible \cite{lj2}. This fact is comforting here, where DOS-related properties are studied.
The substantial gain in computational speed using standard LMTO instead of a full-potential method
is essential when large super cells are to be studied (with a sufficient number of k-points),
since super cells are needed to avoid short range periodicity for disordered cases, and for
studies of dilute impurity concentrations.

T-dependent electronic excitations are included via the Fermi-Dirac function \cite{jar4}. 
The tetrahedron method is used for the k-space integrations, using 343 and 64 k-points in 1/8th of 
the Brillouin
Zone for the small and large unit cell, respectively. Many k-points are needed to describe
the sharp peaks in the
DOS near the gap. Band crossings are not taken into account,
and this explains why the DOS from the two unit cells are not identical for the ordered structure.
The DOS from the 64-atom cell shows some additional 'noise' at some energies, but the gap (and the DOS
near to it)
is identical as it should. The stable behavior of the DOS in the gap region, for varying  
numbers of k-points and of basis size, is important 
because the properties that are studied in this work depend directly on the DOS near the gap.
No calculation of the effective mass from the second derivative
of the band dispersion has been attempted because of the problem with a limited k-point mesh.

Three compositions with doped FeSi were considered; Fe$_4$Si$_3$Al, Fe$_{32}$Si$_{31}$Al and
 Fe$_{31}$IrSi$_{32}$, using one Ir (or Al) impurity atom per Fe (or Si) site in cells of totally
 8 or 64 atoms. The lattice constants were the same as for pure FeSi, and
 no lattice relaxation around the impurity atoms was considered. 

Two methods for the calculation of the exchange enhancement $\cal S$ are employed.   
Spin-polarized calculations are done for the 8-atom cells
including the applied magnetic field, H, which
gives the enhancement ($\cal S$) of the susceptibility. The enhancement is defined as the ratio
between the self-consistent exchange splitting at $E_F$ and the magnetic field energy \cite{jf}. 
This ratio can be different at different k-points and for different bands, while here it
is sufficient to know the Fermi surface average of $\cal S$. 
Since the Fe-d DOS is the dominant DOS-character near the gap (see later) it is convenient
to calculate this average from the exchange splitting of the logarithmic derivative of the 
radial d-wave function 
\cite{jar}.
This value agrees well with the more complicated calculation of the Fermi surface average
of the exchange splitting. This is easily understood from the fact that  
the exchange splitting of the potential,
probed by the d-wave function at $E_F$,  determines the band splittings. In cases with 
disorder there are differences in the exchange splitting of the logarithmic derivative of the d-waves,
so that one can define a local $\cal S$-value. There are smaller variations of the local DOS, 
and the relative variations of local DOS- and $\cal S$-values are quite consistent with the
relation between $\cal S$ and the DOS in eq 1, given below. Further,
it turns out that the exchange splitting of the bands at $E_F$
is close to the Fe-site average of the local $\cal S$-values, as is expected from a perturbative
calculation where the contribution from Si-sites is almost non-existent because of their very low DOS.

The other method to calculate exchange enhancement is via the Stoner factor, $\bar{S}$,
as obtained from  the
para-magnetic bands. Here the method for calculating 
$\bar{S}$ for compounds is used \cite{c15}. The relation between $\cal S$ and $\bar{S}$ is 
\begin{equation}
{\cal S} = 1/(1-\bar{S})
\end{equation}
where $\bar{S}$ is proportional to the DOS, $N$.
This method is based on the assumptions of a spacially uniform exchange splitting, and that 
only kinetic and exchange energies contribute. However, in some situations also the Coulomb energy
contributes \cite{ces}, and the exchange splitting in compounds is in general not uniform in
each atom type, so the calculation of the enhancement  via $\bar{S}$, is less accurate
than via the first method. On the other hand a calculation of $\bar{S}$ is simple, and
only this method could
be applied for the 64 atom calculations, while in the computationally simpler case of 8 atoms/cell
it was possible to compare the results from the two methods. In the following, when we quote
a value of $\cal S$ it is the exchange enhancement calculated by the first method. When it has been
determined via $\bar{S}$ and eq. 1, it is explicitly pointed out. The term "Stoner enhancement"
refers to "exchange enhancement" when using the Stoner model for its estimation.

 The electron-phonon coupling, $\lambda$, is calculated in the Rigid Muffin-Tin
Approximation (RMTA) for compounds  \cite{dac}. This gives the dipole
contribution to the coupling, and even if the DOS at $E_F$ is large in the case
of doping or disorder, $\lambda$ will never be very large. This is because the DOS
is dominated by Fe-d only. For instance, in the regions just below or above the gap, where $E_F$
is located for interesting doping concentrations, 
the percentages of s,p,d and f character on Fe are 1,3,95 and 1 in FeSi,
while in bcc Nb their percentages at $E_F$ are 3,20,73 and 4. This makes  products like 
$N_p \cdot N_d$ and $N_d \cdot N_f$ larger in Nb than in FeSi, even when the total DOS per atom
is similar. Such producs determine the dipolar electronic contribution to $\lambda$, and  
$\lambda$ in doped FeSi will not become as large as in Nb. 
Owing to the weak ionicity of this material, it is not
likely that the monopole contribution to $\lambda$ should be important.

The DOS and Fermi velocitiy, $v_F$, are calculated using the tetrahedron integration method
for the ${\bf k}$-space summation. This permits calculations of the transport
properties such as resistivity, $\rho(T)$, and thermopower, $S(T)$, within Boltzmann theory 
\cite{all}. The resistivity is
the inverse of the conductivity $\sigma(T)$:
\begin{equation}
\sigma(T) = e^2 \tau \int N(E,T) v_F^2(E,T)(-\frac{\partial f(E,E_F,T)}{\partial E})dE
\end{equation}
where $N$ is the DOS, $v_F$ is the Fermi velocity, $f$ the Fermi-Dirac function,
 $e$ is the electron charge and $\tau$ is the scattering life time. $\tau$ is
 assumed to be a constant, and since its amplitude is unknown we calculate 
only relative variations
of $\rho$ due to variations in DOS and $v_F$. In the expression for the thermopower,
$\tau$ is cancelled so that an absolute value can be calculated.
\begin{equation}
S(T) = e \tau T^{-1} \int (E-E_F) N(E,T) v_F^2(E,T)(-\frac{\partial f(E,E_F,T)}{\partial E})dE/\sigma(T)
\end{equation} 
The meaning of the T-dependence in the DOS (and $v_F$)
 will be explained in the next section, where
changes in $N(E)$ due to thermal disorder are considered. In principle, one should
have a smooth, continous evolution of this T-dependence. However,  only a limited number
disordered configurations can be considered because of the large computational effort, and
there will be statistical fluctuations in the T-dependence of 
transport properties and $\chi$, when these properties are calculated from sets of
varying disorders. Therefore, we present often results which are
based on a fixed DOS, and the effects of disorder are discussed qualitatively.

\subsection{Thermal disorder.}

An interesting range of T is when a large part of the phonon spectrum has
been occupied, so that acoustic as well as optical modes are present. The phonon spectrum
can be obtained by a diagonalization of the dynamical matrix for all ${\bf q}$-vectors,
giving phonon energies and phonon wave vectors at all sites. However, the primary interest
for the feedback onto the electronic structure is due to the amplitudes of vibrations
rather than the phonon energies. Therefore, one can proceed more simply and avoid a full
phonon calculation, by considering the vibrational energy of individual sites.

The harmonic approximation considers a parabolic potential well for atomic vibrations.
 The potential
energy $U=0.5 K u^2$ is characterized by a force constant $K$, where $u$  is the displacement,
and the kinetic energy is $0.5 M (du/dt)^2$.  
The sum of time averages of these two energies is constant,
$K u^2$. Putting this equal to the thermal energy $k_BT$, and adding three 
such oscillators (3-dimensions) give
$3 k_BT =  K u^2$, where $u^2 = u_x^2 + u_y^2 + u_z^2$. The relation between
the averaged $u$ and T becomes:

\begin{equation}
  <u^2> = 3 k_B T/ K
\label{eq:K}
\end{equation}
where the force constant $K= M \omega^2$ depends on the averaged phonon spectrum
rather than on the Debye frequency $\omega_D$.
This relation can be obtained from the high-T expansion 
of the proper phonon relations
as long as one characteristic force constant is representative for the vibrations, 
and as long as T is small enough so that the movements of neighboring atoms are
uncorrelated \cite{gri}. 
It is also known that this leads to a Gaussian distribution function
for $u$ with a certain mean-square displacement $ \sigma^2 $ = $< u^2 >$.

\begin{equation}
  g(u)=(\frac{1}{2\pi \sigma^2})^{3/2} e^{-u^2/2 \sigma^2}
\end{equation}

The next term in the high-T expansion for $<u^2>$ is only $\frac{1}{36}$ of the first
term at T equal to the Debye temperature ($\Theta_D$),
 and $\frac{1}{9}$ at $T=\frac{1}{2}\Theta_D$. Therefore equation \ref{eq:K} is a reasonable
approximation even at quite low T, but not at very low T when only acoustic modes are occupied.

Thus for intermediate T we base the estimates of $u$ and $\sigma$ on this simple scheme. 
The force constant is expressed in terms of $\Theta_D$
($K=M (k_B \Theta_D / \hbar)^2$) and $\Theta_D = c \sqrt{a_0 B/M}$, where $c$ is
a structure-dependent constant. 
By taking into account that the calculated  
$B$ is larger than that for most transition metals and assumimg that $c$ is universal,
one can estimate that $K$ is larger than in typical transition metals. From this we
estimate $K$ to be about 10 eV/\AA$^2$. This is consistent with a Debye temperature of
313 K \cite{hun}, and a reduced FeSi mass. An uncertainy of $\sim 20$ \% in $K$ 
is possible and it will
rescale the temperature by the same amount.

Thermal disorder is introduced in the LMTO calculations 
within the unit cells containing 8 or 64 atoms, where 
all  sites become non-equivalent 
as soon as disorder is introduced.
 Each atomic position in the unit cell
is given a random displacement (but totally following the Gaussian distribution) in terms of
 direction and amplitude. The averaged $\sigma^2$
is calculated from all positions and T is defined from eq \ref{eq:K}. Each disordered configuration
represents a "flash-photo" of the dynamical thermal disorder. The assumptions of
Born-Oppenheimer conditions of fast electronic- and slow vibrational- relaxation times, are
likely to be valid, since the lattice stiffness of FeSi is rather normal and since the results
do not depend on very slow electron relaxations. In other words, it is assumed that there is
sufficient time for
self-consistent electronic relaxation during the vibrational evolution in time. The self-consistent
calculations include the 
electronic excitations given by the Fermi-Dirac distribution, but as was shown in ref \cite{jar}
electronic excitations provide only a partial explanation of the T-dependent properties of FeSi. 

Effects of disorder within small unit cells with different disordered configurations 
can be somewhat different in their details,
even if the averaged $<u^2>$ parameter is the same. This is because of the differences
in local configurations, and it can lead to variations of the details in the DOS.
 A more stable DOS at a given T 
can be composed from the
superposition of the DOS from several isothermal configurations, or from the DOS of a much larger
unit cell.  The 64-atom calculations are very slow because of self-consistency, and no spin-polarized
calculations at different T and applied magnetic field  were attemped for that reason.
But since the closing of the gap is similar in the non-magnetic DOS functions from 64-atom and
(configuration averaged) 8-atom cases, one can be confident that the essential physics is described 
within the 8-atom cells.

\section{Results.}

\subsection{Band structure and density-of-states.}

The  bands and the gap of FeSi at T=0 (no disorder) are very similar
to those published earlier using different methods 
\cite{mat}-\cite{ani}. The
calculated lattice constant, $a_0$=4.43  \AA, is found at the minimum of the 
total energy $E_{TOT}$ \cite{jar}.
This is about 1 \% smaller than the experimental lattice constant.
The bulk modulus $B$ at $a_0$ is
calculated to be 2.2 MBars, which is considerably larger than the experimental 
value,  1.3 MBars \cite{kli}. 
 The gap $E_g$ is
indirect, ranging from 6.1 mRy at a lattice constant of 0.98 $a_0$ to
5.7 mRy at 1.02 $a_0$.  This agrees well with Fu {\it et al.} \cite{don} and Mattheiss and Hamann 
\cite{mat} 
who obtained $E_g = 0.11 eV (8 mRy)$ with the ASW
and LAPW band methods, respectively, and with Christensen {\it et al.} \cite{chr} who obtained
0.1 eV using LMTO. The latter work noted only minor changes in the gap structure when so-called
"full-potential" LMTO calculations were performed instead of standard LMTO. 
The fact that the "full-potential" gap
is similar to the other $E_g$, which are based on MT-type potentials \cite{mat,jar,don,ani},
shows that non-spherical potentials corrections are relatively small in the B20 structure.
This fact and the good agreement with the experimental gap are fortunate facts, 
since the computational requirement of supercell calculations with
a full-potental code would have been very large and since the gap value is important
for determining the T-scale for many properties. 
The atomic sphere radii are here 0.328$\cdot a_0$ for Fe and 0.29$\cdot a_0$ for Si. This
gives only small discontinuities of the potential at the limits of the spheres, but
as has been noted earlier \cite{jar,don}, the gap is not sensitive to changes of sphere radii.

It can be noted that calculated band
gaps in typical semi-conductors like Si or GaAs are much underestimated in LDA. In such materials 
the $\ell$-character is different below and above the gap, so that there is a radial change of
states from below to above the gap. As was noted above, the Fe-d character is very dominating
on both sides of the gap in FeSi. It might be that this similarity between an occupied 
ground state below the gap and an excited state above, is essential for a good LDA gap value.
The widening of the gap with pressure is uncommon. This
can be understood from the fact that the gap is in the middle of the Fe-d band. Since the
band width is increasing with applied pressure it is expected that $E_g$ increases as well. 
The value of $d\ln E_g/d\ln V$ is -5.9 from LMTO, which agrees roughly with the observed 
value \cite{gre}.

The DOS functions for ordered and disordered pure FeSi are shown in Fig. 1. The disorder is introduced
as described above in the 64-atom unit cells, corresponding to a certain T. The DOS of
disordered configurations in the smaller cell can be found in refs. \cite{jar3,jar4}. The different
results permit us to conclude
that at a disorder corresponding to about 250 K, the gap is closed but still visible as a dip
in the DOS. At about 450 K no trace of the gap can be seen in the DOS. This process of
gap filling is essential for explaining the properties of FeSi \cite{jar3}. Thermal effects
due to the Fermi-Dirac smearing only, are insufficient to explain the transition of FeSi from
a semi-conductor at low T, to an exchange enhanced para-magnetic metal at $\sim$ 300K.
The DOS on both sides of the gap is dominated by the Fe-d states. The percentage of Fe-d
in the total DOS varies from 85-95 \% in the immediate neighborhood below and above the gap.
For a case with strong disorder the DOS at $E_F$ is about 19 states/FeSi/Ry of which 16 is
from Fe-d. This means that even if there is a sensitivity to variations in T, there are 
no strong charge transfers between Si and Fe sites, nor charge localization within Fe,
as T is changing. However, fluctuations due to differences in local disorder make
the charges to vary among different Fe sites.

Figs. 2-3 show the DOS of pure FeSi,
Fe$_4$Si$_3$Al, Fe$_{32}$Si$_{31}$Al and
 Fe$_{31}$IrSi$_{32}$. The effect of doping is fairly well descibed by rigid-band shifts
 of $E_F$, but some details differ. Replacing a Si with Al appears
 to close the gap more than when an Ir replaces an iron site. The peaks in the DOS on
 both sides of the gap are intact despite doping and $N(E_F)$ changes rapidly with
 small doping concentrations. The derivative of the DOS is so large that it is even 
 difficult to extract very precise values of $N(E_F)$. The Stoner criterion for a
 magnetic instability is roughly situated at 90 states per normal unit cell per Ry,
 so that the peak above the gap is clearly within the magnetic limit, while the peak for
 hole doping is just on the verge of it.

The calculated band filling and broadening of the high peak of the DOS just below the gap
are in good agreement with  optical measurements \cite{sch,deg}. The optical
conductivity showed that the gap, which is clearly visible and in agreement with 
the calculated
gap for low T, is essentially filled up at T $\sim$ 250 K. From the different disordered configurations
we conclude that the complete filling of the gap happens around 300 K. This difference
between theory and optical measurements is quite acceptable in view of the way the disorder
is defined  via an averaged force constant.

The peak below the gap is very narrow, $\sim$ 3 mRy. Angular resolved photo emission (ARPES)
 has been able to identify a narrow band just below the gap, which contributes to this peak
 \cite{par}. It was even possible to observe a T-dependent broadening of this peak.
The DOS peak below the gap is 3-4 mRy wide for the ordered case. This compares well with the
band width for an optimal angle in the high-resolution ARPES of 2-3 mRy, which is observed to
disappear or widen to at least 10 mRy at 275 K \cite{par}. By selecting one angle in ARPES
it is possible to focus on one single band and thereby obtain a smaller band width than
in the DOS. Nevertheless, the peak in the calculated DOS shows a widening at higher T,
which is consistent with the findings of the ARPES data.

\subsection{Magnetic properties.}

First we discuss the results for pure FeSi.
The band broadening and the metallization
are rather quick as T approaches 250-300 K. This leads to an exchange enhancement $\cal S$(T),
and the local $\cal S$(T) (calculated from the ratio between the local spin splitting of the potential 
and the applied magnetic field energy) values are widely spread
on different Fe sites. This can be understood from the spread of local 
disorder \--- at some sites the disorder is strong 
and may cause an attractive Madelung
 shift of the
potential. The Fermi level tends to enter into the high DOS above the (normal) gap
on such sites, while on other sites $E_F$ will be at or below the dip in the DOS. From
this one can understand why the Stoner enhancement, which is sensitive to the
DOS (cf. eq 1), shows local variations from site to site. A striking example of local variations
of DOS and enhancements, are the differences on Fe and Si sites. The DOS in disordered (or doped)
FeSi is typically more than a factor 10 larger on Fe than on Si, and the Si moments in spin-polarized
cases are almost zero or slightly negative \cite{jar}. As a result of this there are large differences
in local exchange splittings.

The magnetic susceptibility $\chi$ can be written as 
\begin{equation}
\chi(T) =\mu_b^2 N_{eff}(E_F,T) {\cal S}(T)
\end{equation}
The effective DOS $N_{eff}$ takes into account electronic excitations
given by the Fermi-Dirac function $f(E,E_F,T)$:
\begin{equation}
N_{eff}(E_F,T) = \int N(E,T) (-\frac{\partial f(E,E_F,T)}{\partial E}) dE
\end{equation}
Here $E_F$ varies with T so that the number of electrons is conserved, and the
band DOS $N(E,T)$ varies with the degree of disorder at the given T, i.e. the DOS is calculated 
for a disordered lattice. 
If $N(E,0)$ is the DOS for T=0, i.e. for a perfect lattice, 
and $\cal S$(T) is 1 (no exchange enhancement), $\chi(T)$
increases
from zero to about 3 $\mu_B^2 / eV$ at 400 K, as in the full line of Fig. 4.
The broken line in Fig. 4 is the result if $N(E)$ is given by the DOS-model of 
ref \cite{man} (two rectangular DOS functions, 5 mRy wide, 880
states/Ry/cell high, and separated by a 6 mRy gap), which is known to reproduce
the experimental results. The peak heights in this model need to be so large since 
no exchange enhancement is included.   Without the effect of DOS
smearing  and exchange enhancement, it is
evident that the  calculated DOS peaks (for T=0) alone
are insufficient to produce the observed susceptibility. As seen
in Fig. 3, the peak heights are of the order 100-200 states/Ry/cell and narrower
than in the DOS model. If now the exchange enhancement is introduced, but still based
on $N(E,0)$, $\chi(T)$ increases from zero
 at 0 K to about 4.5 $\mu_B^2 / eV$ at 400 K \cite{jar}. 
If finally the disorder is considered so that the proper $N(E,T)$-functions are used
(from different 8-atom cell configurations),
 and $\cal S$(T) is an average over the enhancements of all Fe atoms (to give the global
 enhancement as explained earlier), $\chi$ increases
 further, as is shown by the crosses in Fig 4. 
The susceptibility is now quite close to the experimental
 values \cite{jac,man,gre}. 
   The scattering
 of different points reflects the fact that the degree of disorder can be more
 or less pronounced in a small unit cell even if the $u$-parameter is the same.
 
  The local $\cal S$(T) enhancements calculated
 for the cases with largest disorder, varies from 3.5 to 6. This is
not far from the enhancement in fcc Palladium (about 7) \cite{jf,jan}. It 
can be understood from the large local (para-magnetic) DOS values, 
of 11-18 states/Ry/Fe-atom at this large T. This is comparable to the DOS
in many transition metals, like Niobium. The  
very rapid variation of the DOS near the gap in (ordered) FeSi is extraordinary.
Just above the gap it varies from zero to about 50 states/Ry/Fe-atom within 3 mRy. 
In bcc Fe $N(E_F)$ is smaller than this peak value, but clearly sufficient to put Fe
beyond the limit of infinite $\cal S$ \cite{jan}. 
This fact is important for an understanding of the unusual increase of 
$N_{eff}(E,T)$ when T is increased, 
and for the transformation of FeSi from a
semi-conductor at low T to an exchange enhanced para-magnet above room temperature.
If a moderate electron doping is done, so that $E_F$ enters into this
high peak without causing structural instabilities or peak broadening, it seems
possible to have a ferromagnetic state at low T. This possibility is investigated via
the supercell calculations.

It has been proposed that spin-fluctuations are 
responsible for the large magnetic response at large T \cite{mor},
and formation of magnetic moments at high T
has been detected from neutron scattering $\cite{neu}$. This seems plausible
in view of the spacial spread of Stoner enhancement. 
The variation of exchange splittings on different Fe-sites
supports the picture of spin-fluctuations, if the disorder is so large that
the local $\cal S$ tend to diverge at some region at some instance. The time scale of such
fluctuations would be coupled to vibrational time scales, whereas other types of spin-fluctuations
need not to be related to vibrations. The calculations here did not find examples of diverging
$\cal S$ or spontaneous magnetisation coexisting with non-magnetic regions within the same unit cell,
but it could be expected to see such cases for very large disorders.

  The Stoner factors ($\bar{S}$) for the 3 doped cases were calculated to be
0.35, 0.47 and 1.16, for  Fe$_4$Si$_3$Al, Fe$_{32}$Si$_{31}$Al and
 Fe$_{31}$IrSi$_{32}$, respectively.   In the former case, with a small 8-atom unit cell,
 it was possible to calculate $\cal S$ directly and compare the results from the two methods
 described in sect. 2.1.  In this case the Stoner factor (obtained by inverting eq. 1)
 was slightly increased
to 0.43. This shows that not only the exchange and kinetic energies should determine the Stoner factor,
and a full calculation slightly favors the tendency towards magnetism. Still, the calculations
for the larger cells are such that they predict that Fe$_{31}$IrSi$_{32}$, but not
Fe$_{32}$Si$_{31}$Al, should have a magnetic transition. In terms of the DOS, a transition
is likely when $N(E_F)$ per FeSi exceeds about 25 states per Ry. With this limit, it
is for Al-concentrations of 0.07-0.08 when the DOS-peak just
below the gap is close to the required height, that a magnetic transition is most probable
with hole-doped FeSi. The DOS-peak above the gap is higher, making magnetic transitions
more likely for electron doping, if the broadening of the peaks due to strain and disorder
is small as in the calculations shown here.  

The measured magnetic susceptiblity of Fe$_{0.9}$Ir$_{0.1}$Si was presented in Fig. 3 of ref. \cite{sal}.
It has a similar shape to $\chi(T)$ in pure FeSi. However, those data had been corrected
for an assumed para-magnetic iron component corresponding to 1.5 \% Fe, and the raw data
were quoted as having a minimum at 120 K, and the amplitude of this minimum was comparable with
the maximum of the susceptibility of pure FeSi. Unfortunately
no other details are given concerning the uncorrected data. The calculations
produce an almost magnetic state for this Ir concentration. Rigid bands, using both
the FeSi and Fe$_{31}$IrSi$_{32}$ DOS, puts the shifted $E_F$ just on the right hand side of
the high peak above the gap, above or near the Stoner criterion for magnetic order at T=0.

A calculation of the susceptibilty needs to take into account both Stoner enhancement and
thermal disorder, as was concluded for pure FeSi. In the doped case this will be difficult, since
the magnetic ordering, expressed by a diverging $\cal S$(T), is so close, and spin-polarized
calculations for several disordered configurations are too cumbersome. Some insight is provided
by the evolution of $N_{eff}(E_F,T)$. Fig. 5 shows this quantity for
Fe$_{0.9}$Ir$_{0.1}$Si as function
of T, using rigid-band shifts on the DOS from various disordered cases. No minimum as a function of T
is found when the same DOS is used over the whole T-range. But when the disorder is increased,
it first diminishes the DOS and than increases it again, so when $N_{eff}(E_F,T)$ is respresented by
panels form various different disordered cases, as is shown by the crosses in Fig. 5,
a minimum emerges somewhere around 200-300 K. The scattering of points is a result of the statistical
fluctuations of disorder in small cells as mentioned earlier. The appearance of a minimum can
be understood from the position of $E_F$ on the edge of the peak for this composition, so that smearing
will tend to increase the effective DOS only at the highest T. The Stoner factor will enhance
the structures further.
The tendency towards magnetic ordering is even more pronounced at intermediate Ir concentrations;
at x=0.04-0.05, just above the concentration of the 64-atom unit cell, when the DOS has its maximum. 
It is therefore likely that the raw data mentioned in ref \cite{sal} represent the intrinsic 
behavior of Fe$_{0.9}$Ir$_{0.1}$Si.

 One spin-polarized calculation was done for
the Ir doped, ordered supercell, Fe$_{31}$IrSi$_{32}$. The self-consistent calculations 
are very slow and only few k-points could be used. The partially converged results
indicate that the total magnetic moment is able to reach the 1 $\mu_B$ that is needed to make the
supercell semi-metallic for this composition, so that $E_F$ falls in the gap for minority
spins. This is at low T when the gap still is sharp. The moment is of the order 0.03 $\mu_B$ per Fe-atom.
 This moment seems very small in view of the reasonably large 
Stoner factor, but it can be understood from
the fact that the peak in the DOS is so narrow. At larger T, tails of the DOS will enter
 the gap and both the majority 
and minority Fermi levels are within finite DOS values. The moment can remain for some doping
concentrations as long as the peak in the DOS is large.

The local variations of the magnetic moment in the Ir-doped supercell show a clear trend. Near
the impurity the Fe-moments are largest, while at the most distant Fe sites, the moment is only about a
third as large. Similarly, the local Fe-DOS values in the non-polarized results are highest
near the Ir site, while far from it the Fe-DOS values are smallest. The trend for an even larger
super cell goes in the direction of having an electronic structure like undoped FeSi far from
the impurity site, so that only the region near the impurity would become polarized. As was mentioned
above, the relative variations of the local moments are larger than for the DOS, which is consistent
with the relation between DOS, $\bar{S}$ and $\cal S$, given by eq. 1. Further, the similarity of having local 
variations of the DOS in disordered pure FeSi and in doped FeSi, supports the idea that spin-fluctuations
can be coupled to vibrational disorder. However, with the cases of general disorder studied here, the
local variations have no long-range correlations as in the case with the dilute impurity, and the degree
of disorder is limited so that no cases of spontaneous magnetisation are found.

Strong ferromagnetism is found in other transition metal silicides of the same B20 structure.
MnSi, for instance, has large magnetic moments in agreement with spin-polarized band calculations
\cite{lj}, and the paramagnetic DOS on Mn is large to make the value of $\bar{S}$ larger than one.

Finally a comment should be made about the possibility of a metamagnetic transition. It has been argued that
metamagnetism can occur as a result of strong correlation \cite{ani}. This is plausible
already from the LDA results, if an applied field is able to close the gap at low T, when the DOS peaks are 
sharpest \cite{jar}.  The problem is that the applied
field has to be very strong and that the magnetic state goes away if the field is removed,
at least according to the LDA calculations. 
Thus the total energy seems to be lowest for the non-magnetic
case, with no barrier towards a magnetic one. However, magnetic impurities can provide a field
via hybridization, and this might explain the upturn in $\chi(T)$ at very low T, that is found
in most FeSi samples. The case of Ir doping as mentioned above is one exemple of impurity induced extended magnetism,
when the impurity play the role of electron donor. Magnetic impurities with localized states
act differently via their exchange splitting, and the magnetisation is likely to be localized around the
impurity because of the short reach of hybridization. At low T the role of localized states in impurities will 
certainly influence the properties of otherwise undoped FeSi. This has been investigated experimentally
by Hunt {\it et al.} \cite{hun}.

\subsection{Heat capacity}

The electronic specific heat is linear in T; 
$C_e = \gamma T = \frac{\pi^2}{3} k_B^2 T N_{eff}(E_F,T) (1+\lambda)$, as long as the DOS has a weak
T-dependence.
A comparison of the specific heat $C_p$ 
between FeSi and CoSi revealed an "anomaly" in FeSi
of about 3-5 J/mol/K \cite{jac,man}, above about 200 K. This contains a great
deal of uncertainty, since the extraction of the anomaly is based on the 
assumption of
identical lattice contributions to $C_p$ in the two compounds.

The anomaly with respect to CoSi is calculated from the (electronic) 
kinetic energy, ${\cal U}$, as in the model of ref \cite{man}. If  $\lambda$ is ignored;
\begin{equation}
{\cal U} = \int N(E) E f(E,E_F,T) dE
\end{equation}
and $\Delta C(T) = \frac{d}{dT}{\cal U}$. (The DOS of ordered FeSi is used, because of
the problem of connecting results from different disordered configurations.)
The results from the DOS of the small and large cell,
are shown in Fig. 6 together with a model made to resemble the experimental data
in ref \cite{man}, although the model parameters are different from the ones in that reference. 
The results follow the model quite well with a maximum not far from  $\sim 300$K,
where the measured $\Delta C_p$-anomaly has its maximum.  
At large T it is expected that $\lambda$ should decrease \cite{gri}, but the calculated
$\lambda$ is not large, and this should concern the reference system as well. A possible
complication is that $\gamma T$ for the reference system, CoSi, may show a strong non-linear
evolution at high T. Indeed, the calculated DOS for CoSi shows that $E_F$ is situated right at the 
DOS shoulder about 40 mRy above the gap, and that the effective DOS tends to increase
when the DOS is broadened. However, this is only a qualitative observation; calculations
of differences in $C_p$ is difficult and the agreement with experiment shown in Fig. 6
is surprisingly good. If the DOS from a disordered configuration is used, it will increase
the heat capacity more at low T, while at larger T the result is not so different from
what is shown in Fig. 6.

It is worth noting that the DOS model made by Mandrus {\it et al.} to fit $\chi(T)$-data
required higher and wider DOS peaks than the model made to fit $\Delta C_p(T)$ \cite{man}. 
The spectral weights in the DOS were 4.4 and 0.8 electrons, respectively. The difference can be
interpreted as if an exchange
enhancement of about 5 is included only in $\chi$.
This is consistent with our finding which has an enhancement of this order  
(cf. Fig. 4) in $\chi$,
but is absent in the heat capacity.  

The low-T specific heat of doped FeSi$_{1-x}$Al$_x$ for $x=0.025$ and 0.015 was
measured by DiTusa {\it et al.} \cite{ditusa}. They noted a large effective mass of about 14
for the largest doping concentration. This was concluded from the large $\gamma$ of
about 7 and 5 $mJ/mol Fe K^2$, for the two concentrations. The calculation for the Fe$_{32}$Si$_{31}$Al
supercell has a DOS of about 700 states per cell Ry, and $\lambda$ is calculated in RMTA 
to be 0.2 for a Debye
temperature of 313 K, taken from ref \cite{hun}. Rigid band shifts of $E_F$ on the DOS from pure 64 or 8 atom FeSi calculations
do not change these estimates much. This gives $\gamma = 4.6 mJ/mol Fe K^2$ , slighly smaller
than found experimentally. Enhancement from spin-fluctuations could make up for the difference
since $\bar{S}$ is quite large. If $E_F$ is adjusted in a rigid-band manner to account for
$x=0.015$ one finds a reduction of $\gamma$ to about 3 $mJ/mol Fe K^2$. Thus the relative change in $\gamma$
between the two concentrations is reflected in the sharp change in DOS between the two cases.
The different amplitudes in experiment and theory can be translated into a 30-40 \% lower $m^*$ in
the theory. Compared to doping of conventional semi-conductors, FeSi is exceptional because
of the high DOS peaks at the gap edges. This is reflected in the large band mass of doped FeSi.

\subsection{Transport properties.}

The thermopower, expressed by the Seebeck constant, $S(T)$, is
 calculated from the bandstructure of ordered (very slightly
hole doped) FeSi \cite{jar}. It agrees
very well with experimental data of Sales {\it et al.} \cite{sal}, both concerning amplitudes
and T-dependence up to 100-150 K \cite{jar}. The experiment is for supposedly pure FeSi,
but the agreement with the calculated $S(T)$ shown in Fig. 7 suggests
that there is a slight hole doping. A slight hole doping (which could even be smaller than 0.005 holes
per cell) is needed to make a shift in $E_F(T)$ from near the top of the valence band at low T,
to a little below the middle of the gap at high T. Exact stoichiometry implies a smaller
shift of $E_F(T)$, since $E_F$ at T=0 would be just at the middle of the gap, and the amplitude
variation of $S(T)$ would be smaller. Doping with electrons makes a large difference,
since $E_F(T)$ then shifts downwards from the edge of the conduction band to just below the mid-gap,
 as T increases, and $S(T)$ takes an opposite shape, as shown in Fig 7.
 
 The good agreement between experiment and theory for $S(T)$ for T up to $\sim$ 150 K, 
 gives an indication that effects 
of vibrational
disorder are not very important up to this temperature, at least not for pure FeSi. $S(T)$ is small and flat for
higher T, and calculations  cannot distinguish clearly
between different disorders.

Experimental data for Ir-doped FeSi with x=0.03 show a completely different $S(T)$ \cite{sal},
as is expected from an electron doped case.
The calculated result, shown in Fig 8, using the (ordered) FeSi band with $E_F$ moved to account for the electron
doping at x=0.03, shows a very similar shape as in the measured curve in ref \cite{sal},
 with a minimum of $\sim 0.15 mV/K$, 
except for the fact that the temperature
scale is different by a factor 3 roughly. The dip in $S(T)$ is near 300 K, while near
100 experimentally. Qualitatively, it is expected that additional DOS smearing above
100-150 K, will compress the T-scale, to agree better with experiment. But is is
difficult to put together results from several band structures with different T-dependent
disorder, to make a continous, quantitative plot of the thermopower. Nevertheless, in Fig. 8 
is shown the result of $S(T)$ for x=0.1, when it is collected from several
different disordered configurations, each one representative of the corresponding
 interval in T. Despite
the scattering of points, it is possible to identify a minimum near 300 K, which is not
found in the curve based on the T=0 bands only (the broken line). The amplitude agrees 
well with experiment, but the minimum is still at too high T. A possible explanation of this is that
non rigid-band effects and spin-polarization will modify the peak structures at this relatively high doping 
concentration.

A similar trend concerning the peak positions, 
is obtained for the resistivity $\rho(T)$ in Fe$_{1-x}$Ir$_x$Si for x=0.03, 
cf. Fig. 9.
The experimental peak is found near 90 K \cite{sal}, while it is near 300 K from the band results. 
For x=0.1 $\rho(T)$ is increasing with T, both in the experiment and the calculation
(assuming a constant life time $\tau$), with no maximum within a wide T-range.

Measurements of the conductivity of hole doped FeSi$_{1-x}$Al$_x$ with x=0.1 has been
published by Sales {\it et al.} \cite{sal}, and recently within the concentration range x=[0-0.08], 
by DiTusa {\it et al.} \cite{ditusa}. The conductivity has a minimum in the T-range 50-100 K
when x exceeds $\sim 0.05$. The calculated conductivity shows the minimum, but again as in
the case with electron doping, the peak position is displaced towards higher T, when the 
calculations are based on rigid-band shifts of $E_F$ on the bands
of an ordered structure. An example is shown in Fig. 10.

Finally, Fig. 11 shows the evolution of $S(T)$ in Al-doped FeSi, with and without considerations of 
thermal disorder. As before, the T-scale is compressed by disorder, while the experimental peak 
is found at even lower T for the Al-concentration x=0.1 \cite{sal}.

Thus, pure FeSi shows very good agreement with experiment, while in the
doped cases, the T-scale is somewhat renormalized due to disorder. This can be understood
from the fact that $E_F$ is within the DOS at all T when the doping is important.
In pure FeSi, the position of $E_F$ is in the gap and it it is not until a certain T that
the band edges come close to it.
Therefore, as long as $E_F$ is within the gap region, the properties are not so sensitive
to broadening due to disorder. With doping, $E_F$ is within the DOS even at low T, and smearing
has an immediate effect on the properties. This often show up as a difference in T-scale in 
measured and calculated transport properties.
 On the other hand, the calculated amplitudes of $S(T)$ are always in
good agreement with experiment.

\section{Discussion and summary}

In conclusion, we have performed non-magnetic as well as spin-polarized band
calculations for ordered and disordered FeSi with different dopings. Various 
properties are calculated from the
band results, and the agreement with experiment is, in general, very good.
Most importantly, it has been possible to show that the unusual behavior of the susceptibility
can be described within LDA.
Structural disorder, which can
be linked to the temperature, leads to
a gradual closing of the narrow gap, so that the effective DOS 
(per Fe) at the Fermi energy
becomes comparable to that of Nb or Pd, for T larger than $\sim$400 K. The increased DOS
has a direct impact on the Pauli susceptibilty, and on the
exchange enhancement. By taking these effects into account
one obtains a $\chi(T)$ which is of the correct amplitude. This is true for the heat capacity
as well, where no Stoner enhancement enters, and the calculation based on the normal DOS
reproduces the essential part of the anomaly. The direct observation of bands and
DOS in photoemission and optical experiments are consistent with the calculated
electronic structure when disorder is considered. 
The transport properties are well described, except
for what seems to be a somewhat too extended T-scale in the case of doped FeSi.
The thermopower in pure FeSi shows that thermal disorder
is not crucial below $\sim$ 150 K, for this undoped case. 

Thus, as a whole we obtain a very satisfactory explanation of the properties of FeSi
from LDA, provided that thermal disorder is taken into account. The band calculations are ab-initio so
the only uncertainty concerns the exact form of disorder. Despite
the simplicity
of the harmonic model for vibrational disorder, the band results lead even to a good 
quantitative description of many properties. Future work should improve the calculations of 
the phonon part, and larger unit cells
and/or more disordered configurations are required to achieve less scatter
in the calculated results. But such elaborations are not expected to modify the qualitative 
results very much for intermediate T. 
The exact behavior at the lowest T, which is not studied here, is delicate because of the 
relative importance of zero-point motion and low-q phonons.
Different types of disorder have some influence on
the results, but it is especially the difference between having disorder or not,
which is most  important in this work.
 Thermal disorder is always present above a certain temperature,
and the consequencies are particularily striking in FeSi because of its sharp DOS 
features near the narrow gap.
These results imply that no assumptions of strong correlation are needed to explain the
properties of FeSi.

\vspace{0.5cm}
{\noindent {\bf Acknowledgements:}}

I am grateful to S. Dugdale for a careful reading of the manuscript and to J.F. DiTusa for sending
the preprint in ref \cite{ditusa}. 

\newpage
\noindent
{\bf Figure Captions}

\begin{enumerate}
\item
Paramagnetic density-of-states of FeSi (64 atoms per cell) without disorder
(thin line), with disorder corresponding to a temperature of about 250 K (broken line),
and with disorder of about 450 K (bold line). 
The energy is relative to $E_F$.

\item
Paramagnetic density-of-states of ordered Fe$_{32}$Si$_{32}$ (thin line), Fe$_{31}$IrSi$_{32}$ 
(bold line) and Fe$_{32}$Si$_{31}$Al
(broken line). 
The energy is relative to the gap center, so that the positions of $E_F$ are indicated by
the vertical broken line for Fe$_{32}$Si$_{31}$Al, by the vertical line for Fe$_{31}$IrSi$_{32}$
, and is at zero energy for Fe$_{32}$Si$_{32}$.

\item
Paramagnetic density-of-states of ordered Fe$_4$Si$_4$ (thin line) and Fe$_4$Si$_3$Al (bold line). 
The two curves are lined up relative to the respective position of $E_F$.

\item
Magnetic susceptibility $\chi$ for FeSi. The full line is calculated from the ordered DOS
with electronic excitations only. The broken line is the result using
the DOS-model of ref \cite{man}, which closely 
reproduces the experimental susceptibility. Calculations using the DOS from nine different
disordered configurations (corresponding to different T) using 8-atom cells, 
but without the $S(T)$-factors, 
are shown as circles. The results including the exchange enhancements $S(T)$ from
spin-polarized calculations are shown as crosses.

\item
Effective DOS at $E_F$ for Fe$_{0.9}$Ir$_{0.1}$Si obtained from shifting $E_F$ to account
for the Ir doping. The full and broken lines are the result using ordered and disordered
(about 550 K) structures, respectively. By using panels corresponding to different 
intermediate disorders one obtains the crosses, where vaguely a minimum can found somewhere
near
200-300 K. The effective DOS times the Stoner enhancement describes the susceptibility $\chi(T)$.
Experiments show a minimum of  $\chi(T)$ near 120 K \cite{sal}.

\item
Specific heat calculated from the band results for ordered structures as described in
the text. The result using the DOS from the small cell is shown by the circles, and 
from the 64 atom cell by the crosses. The results are somewhat unstable at low T. The
full line is from a DOS model made to reproduce the experimental points for $\Delta C_p(T)$
shown in ref \cite{man}.

\item
The calculated Seebeck coefficient for (ordered) FeSi, where $E_F$ has been adjusted for
slight hole and electron doping, respectively. The good agreement between the hole doped case
and the experimental data in ref \cite{sal}, suggests that effects of thermal disorder
are minor up to 150-200 K.

\item
The calculated Seebeck coefficient for ordered and disordered Fe$_{0.9}$Ir$_{0.1}$Si.
The band structure is for the 8-atom cell, with $E_F$ shifted to account for the Ir-doping. 
The broken line is for the ordered stucture. The crosses show $S(T)$ when it is
made up from different panels with different degree of disorder. 
Compared to the ordered case, it is seen how the minimum is displaced to
lower T.  
The full line is for the ordered Fe$_{31}$IrSi$_{32}$.

\item
Relative evolution of the resistivity in Fe$_{31}$Ir$Si_{32}$ and Fe$_{0.9}$Ir$_{0.1}$Si.
In the latter case $E_F$ is shifted to account for the increased Ir-doping. Experimental
data in ref \cite{sal} show a resistivity peak at lower T.

\item
Relative evolution of the conductivity in FeSi$_{0.9}$Al$_{0.1}$ for ordered structure
(bold line), and with disorder of about 250 K (thin line). The latter agrees with the
experimental data in ref \cite{ditusa}, which show a conductivity minimum at about 100 K.
The minimum tends to lower T, and give lower conductivity when $E_F$ is moved closer to
the gap by accounting for 0.019 (broken line) and 0.006 (dotted) holes per FeSi. In the case
of ordered structure, $\sigma$ tends to zero at T=0 when FeSi is undoped.

\item
The calculated Seebeck coefficient for ordered and disordered FeSi$_{0.9}$Al$_{0.1}$.
The band structure for the 64-atom supercell containing one Al atom was used with
$E_F$ adjusted to account for the higher Al-concentration. It is seen that the effect
of disorder is to compress the T-scale.

\end{enumerate}

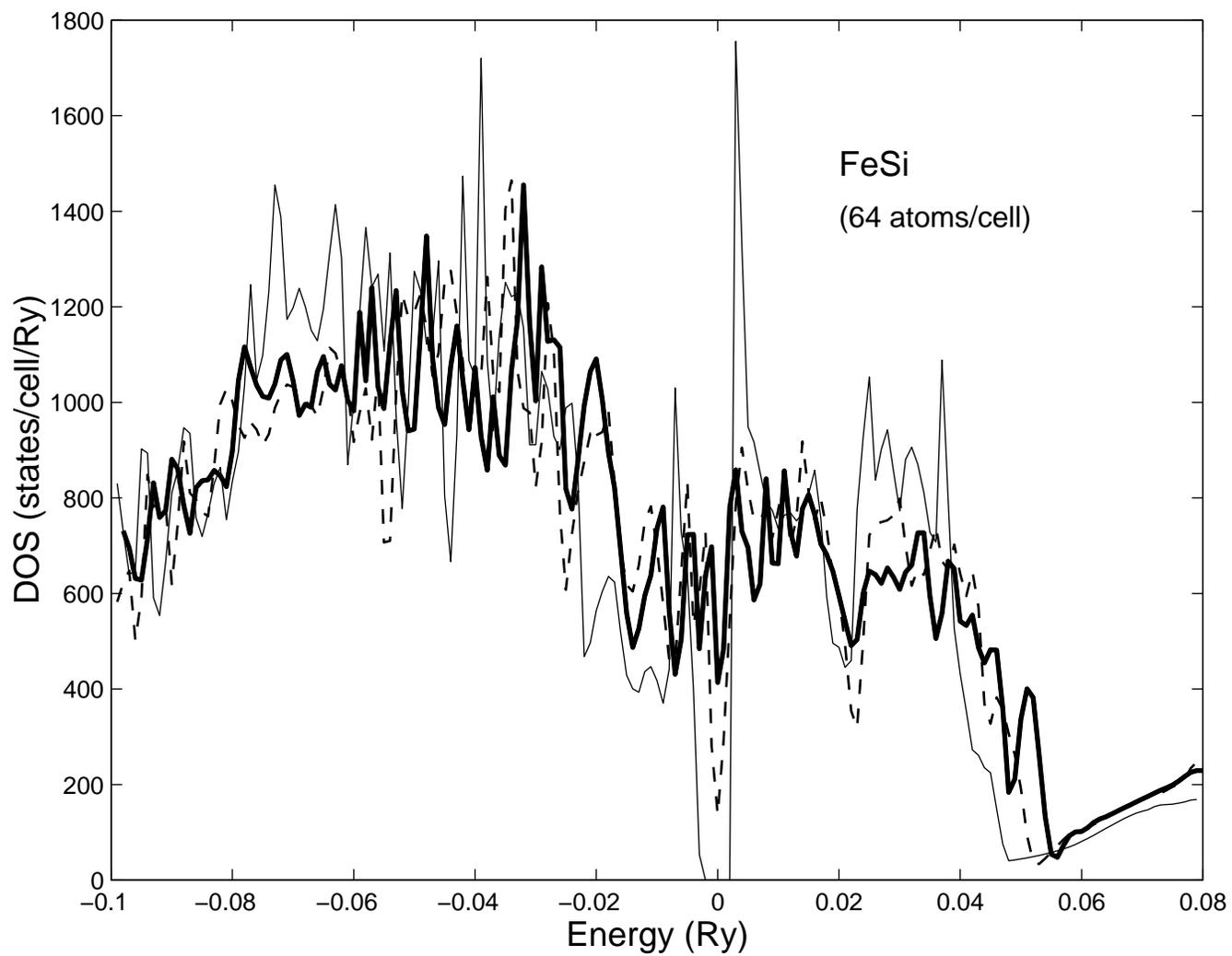
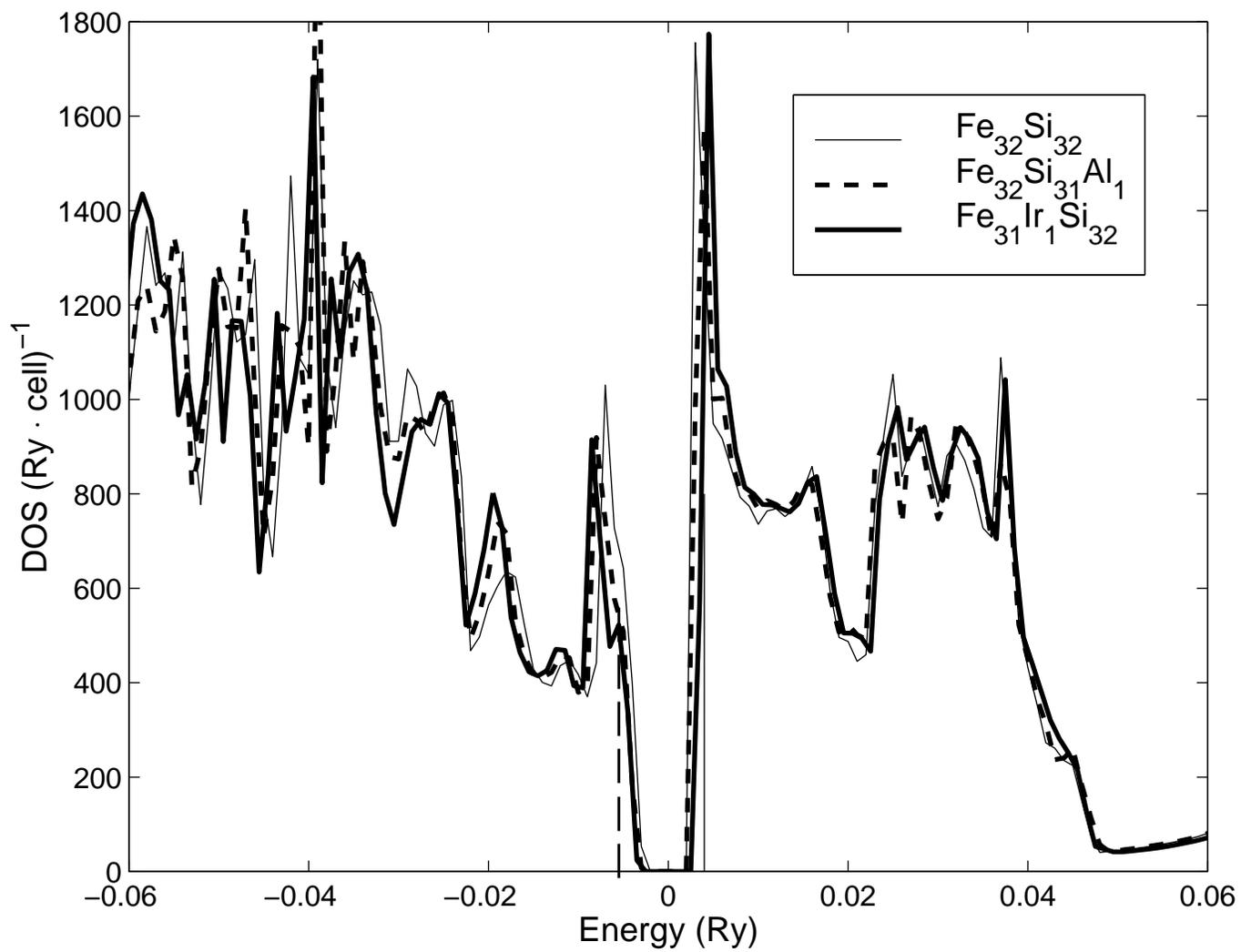
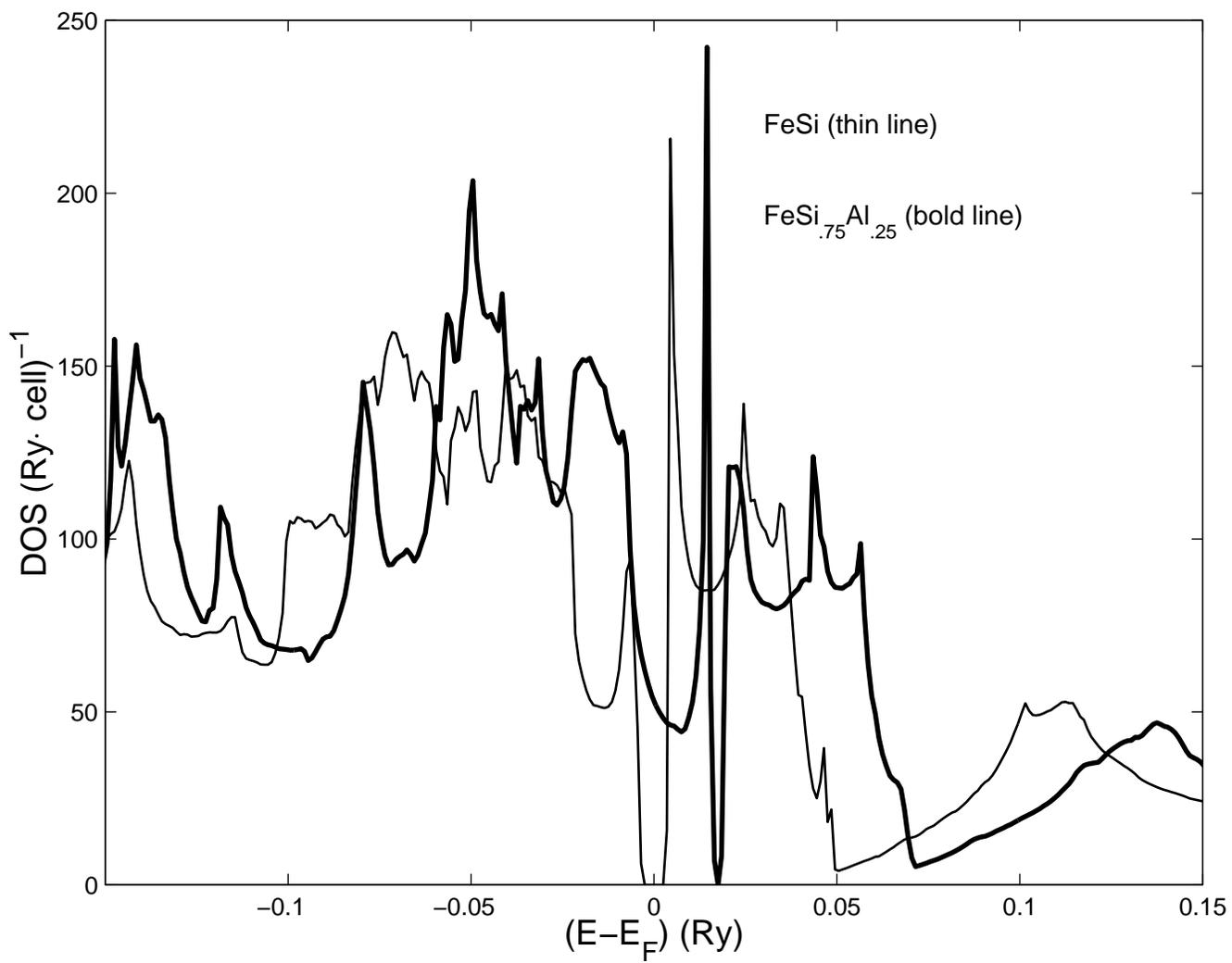
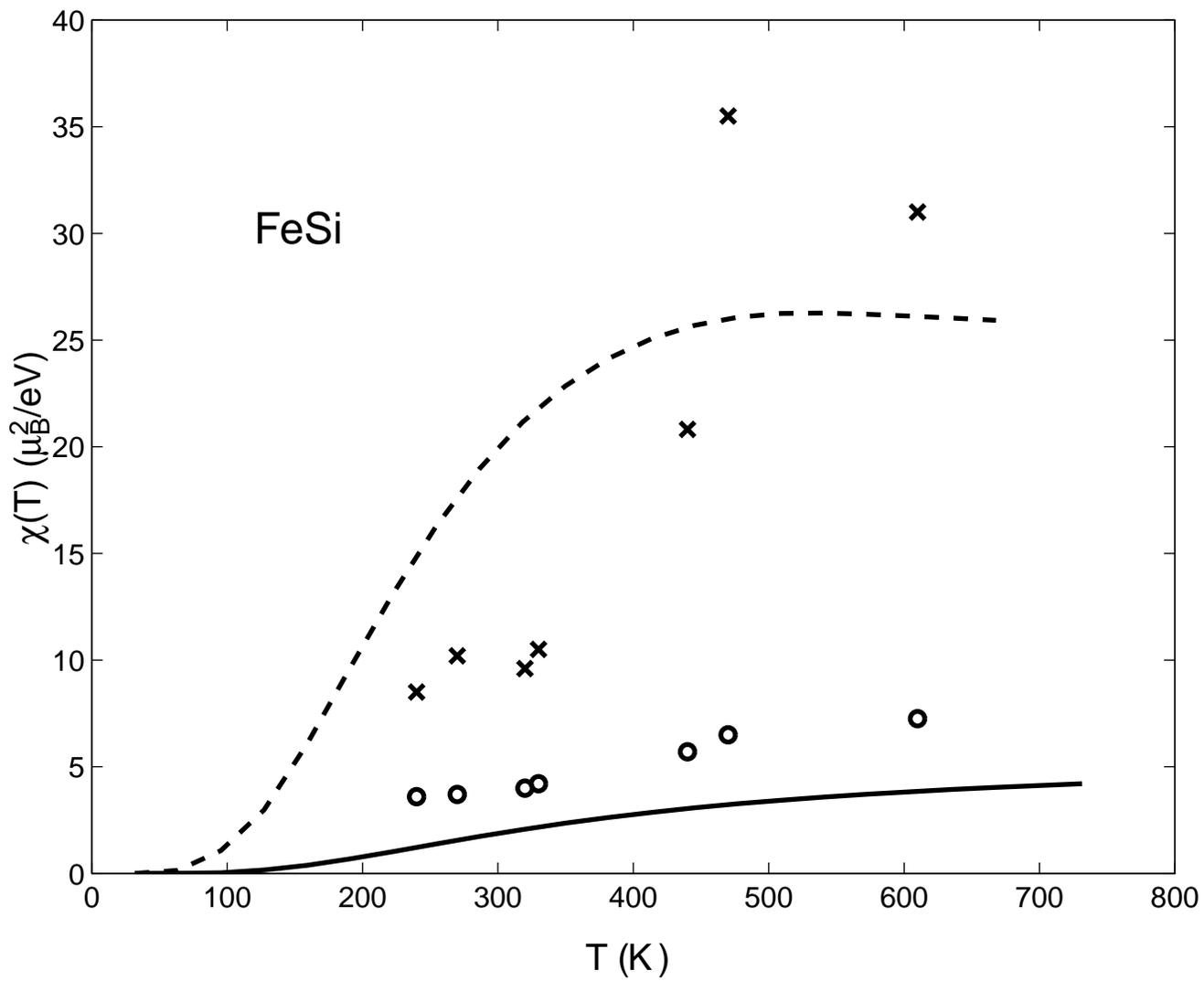
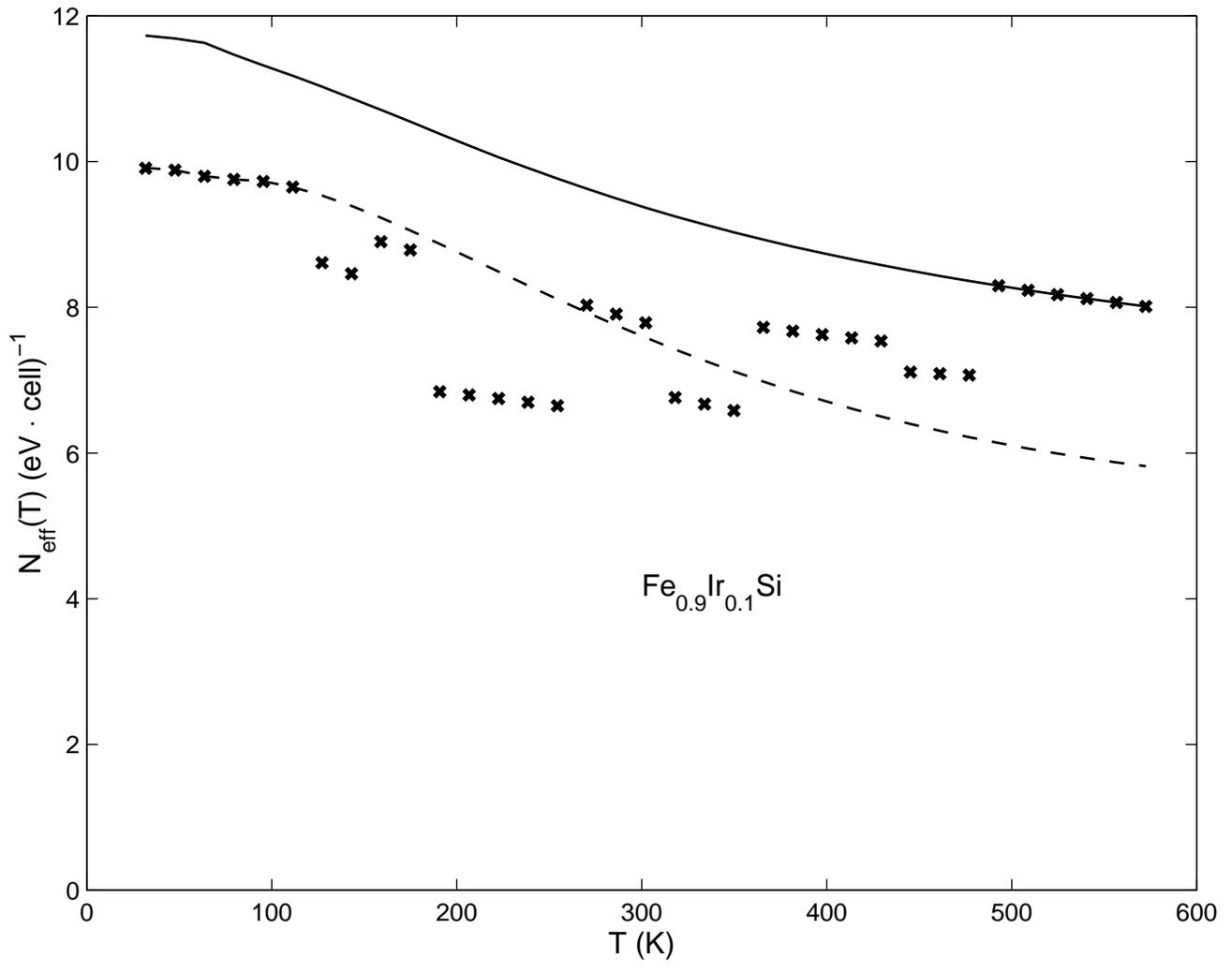
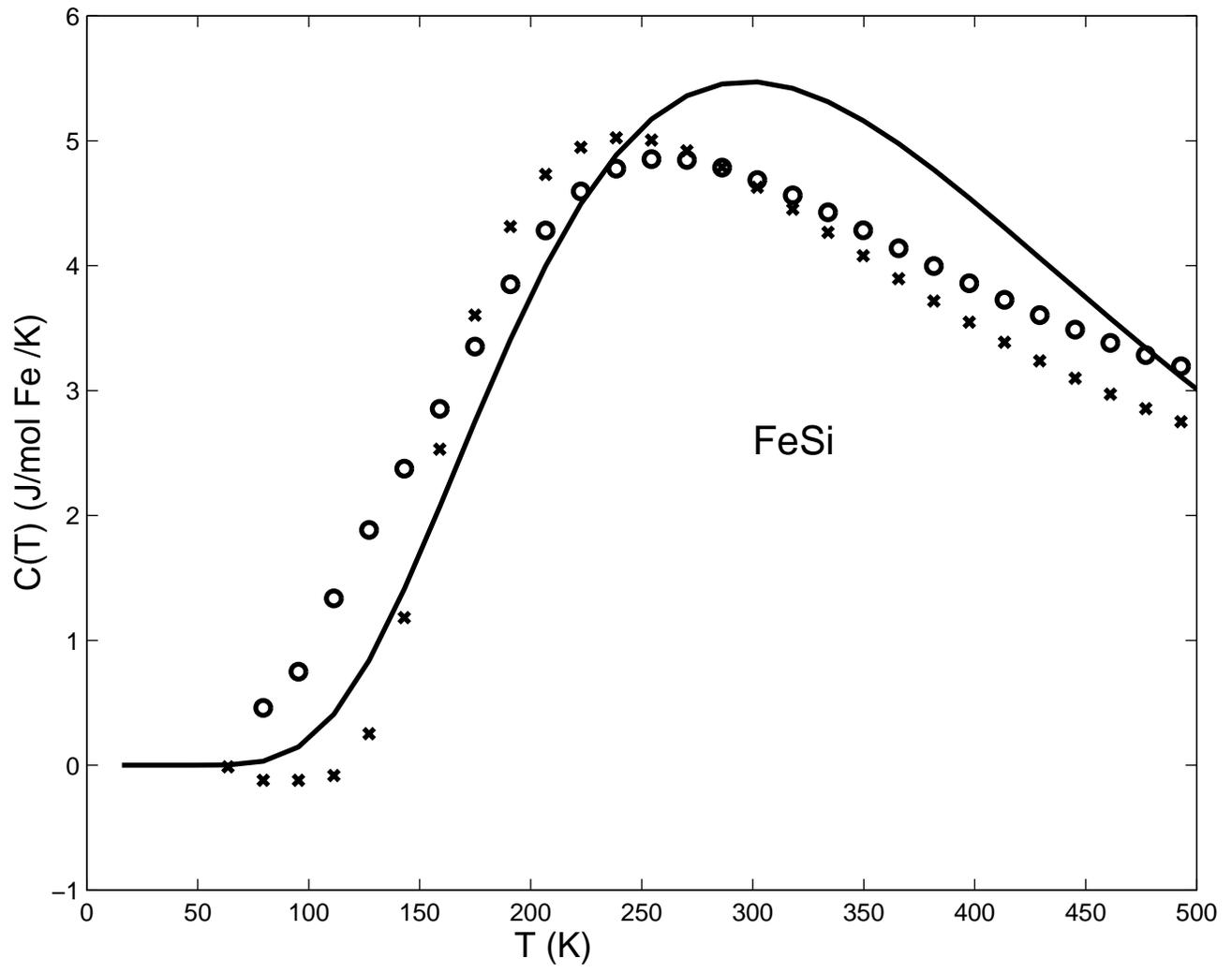
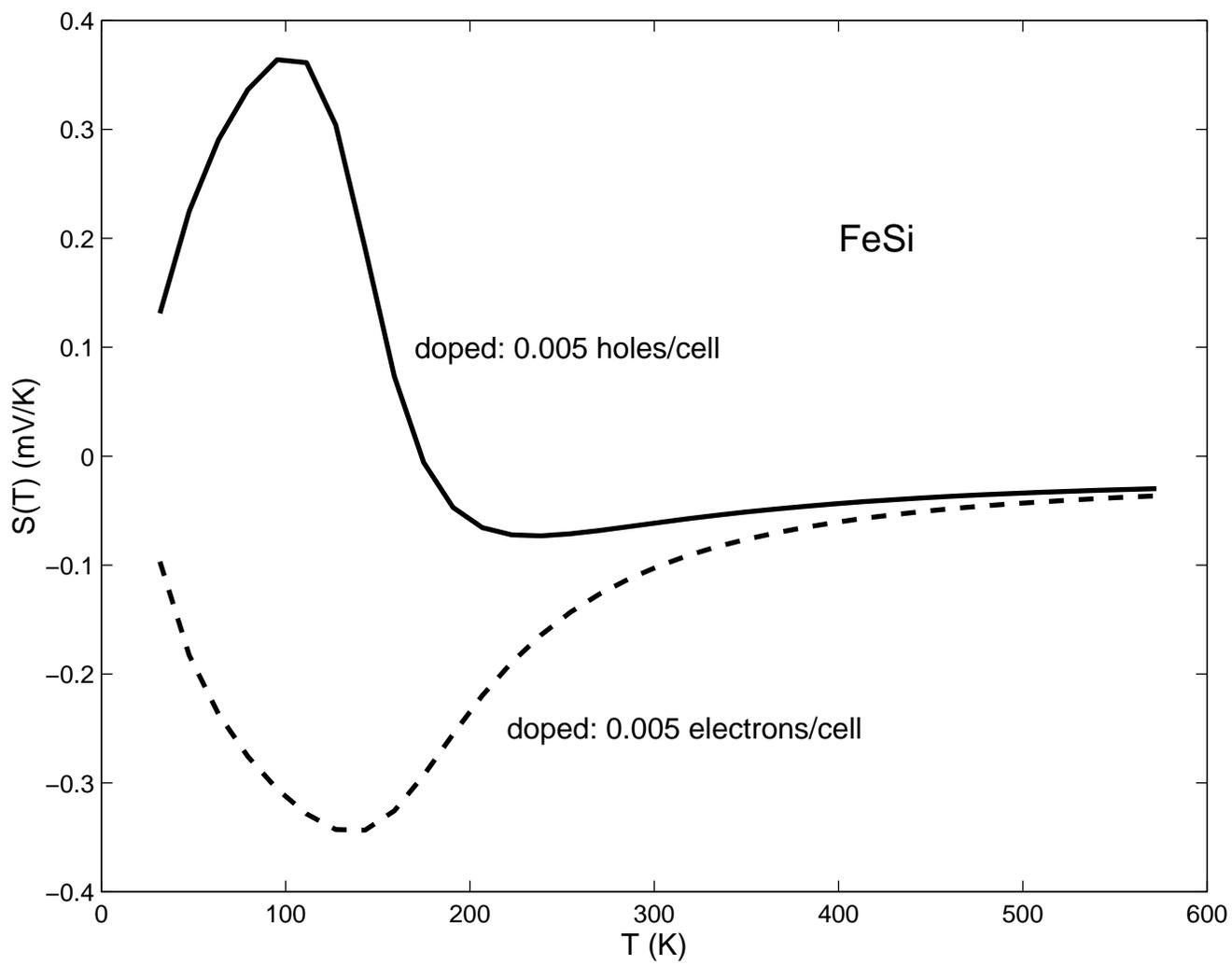
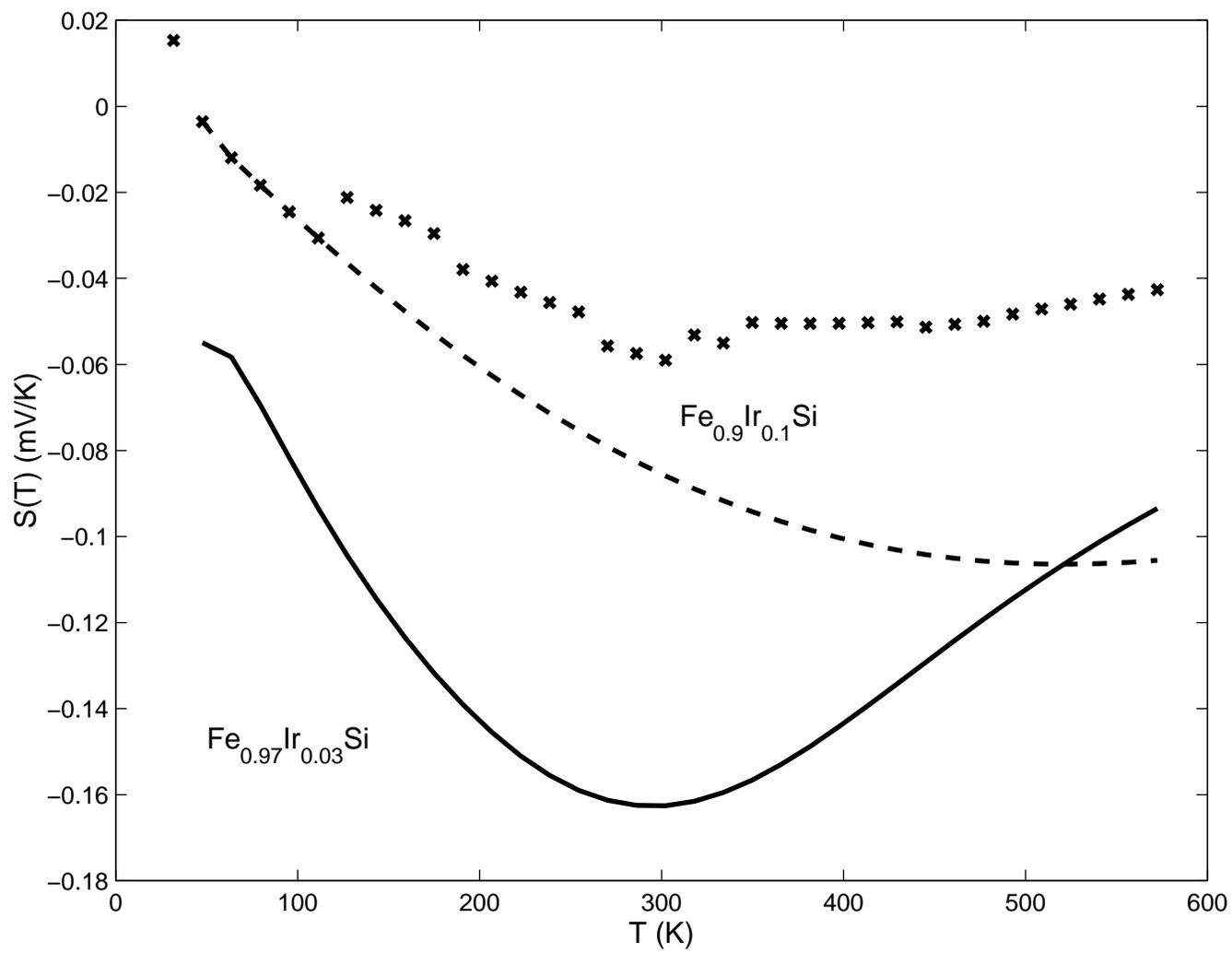
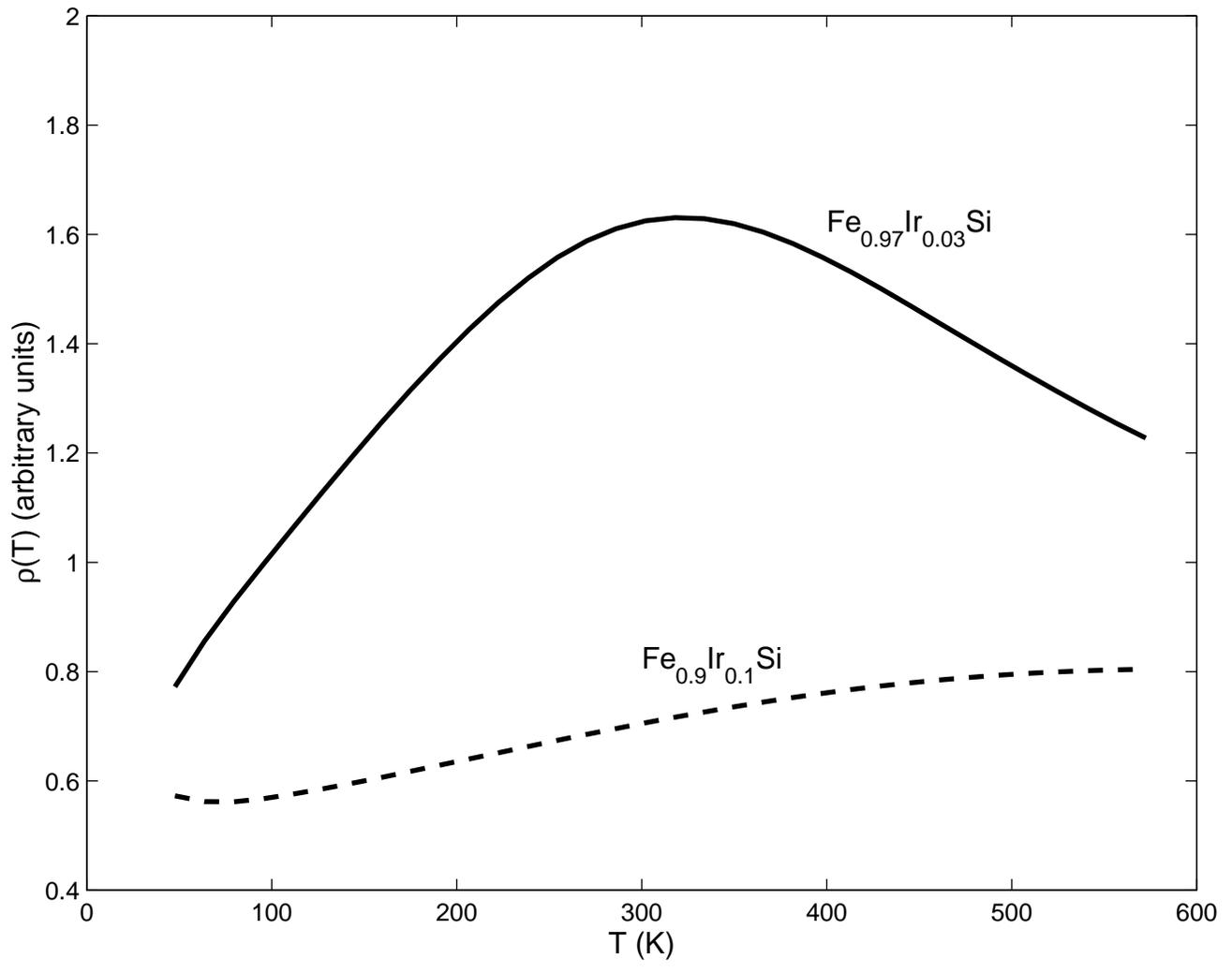
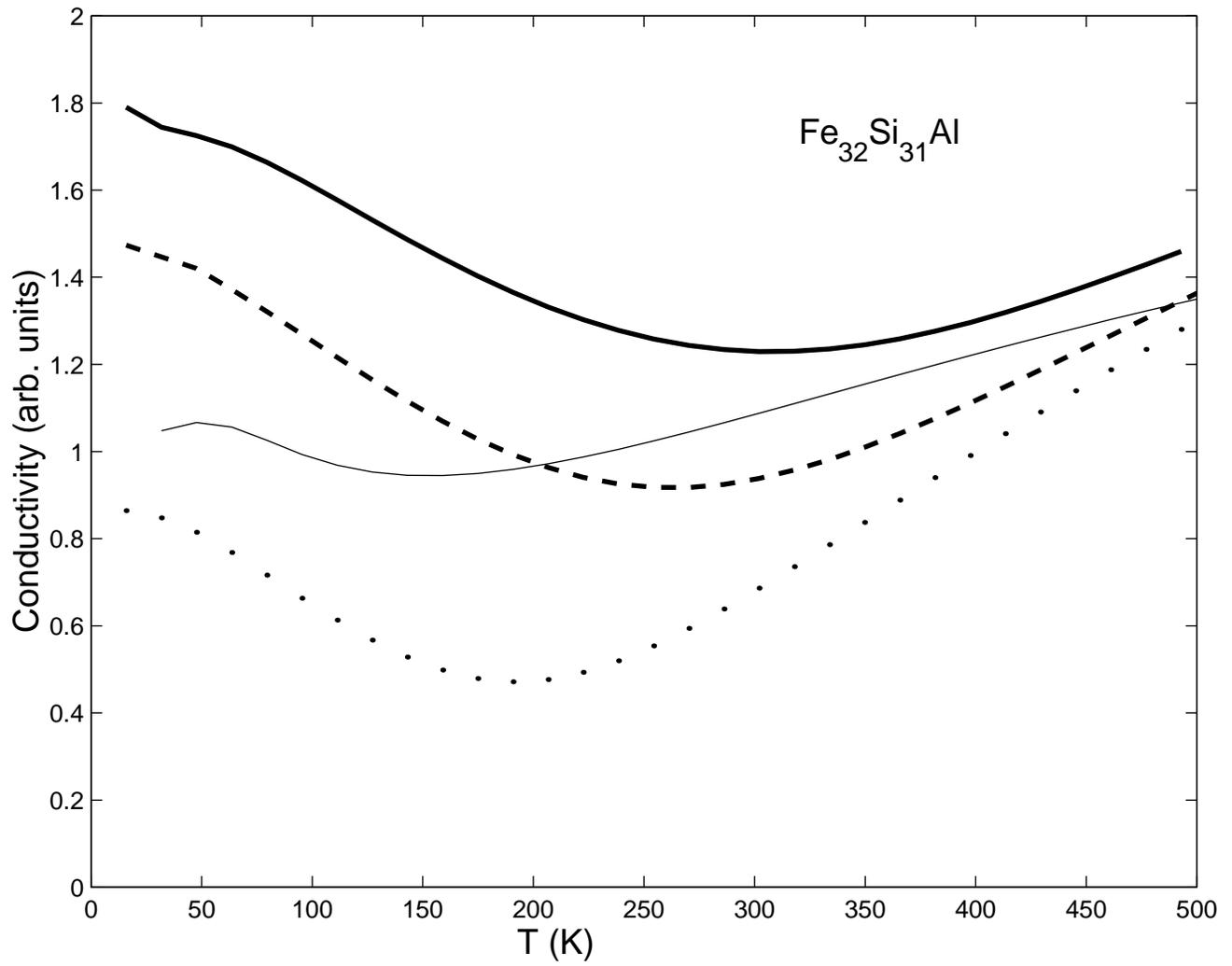
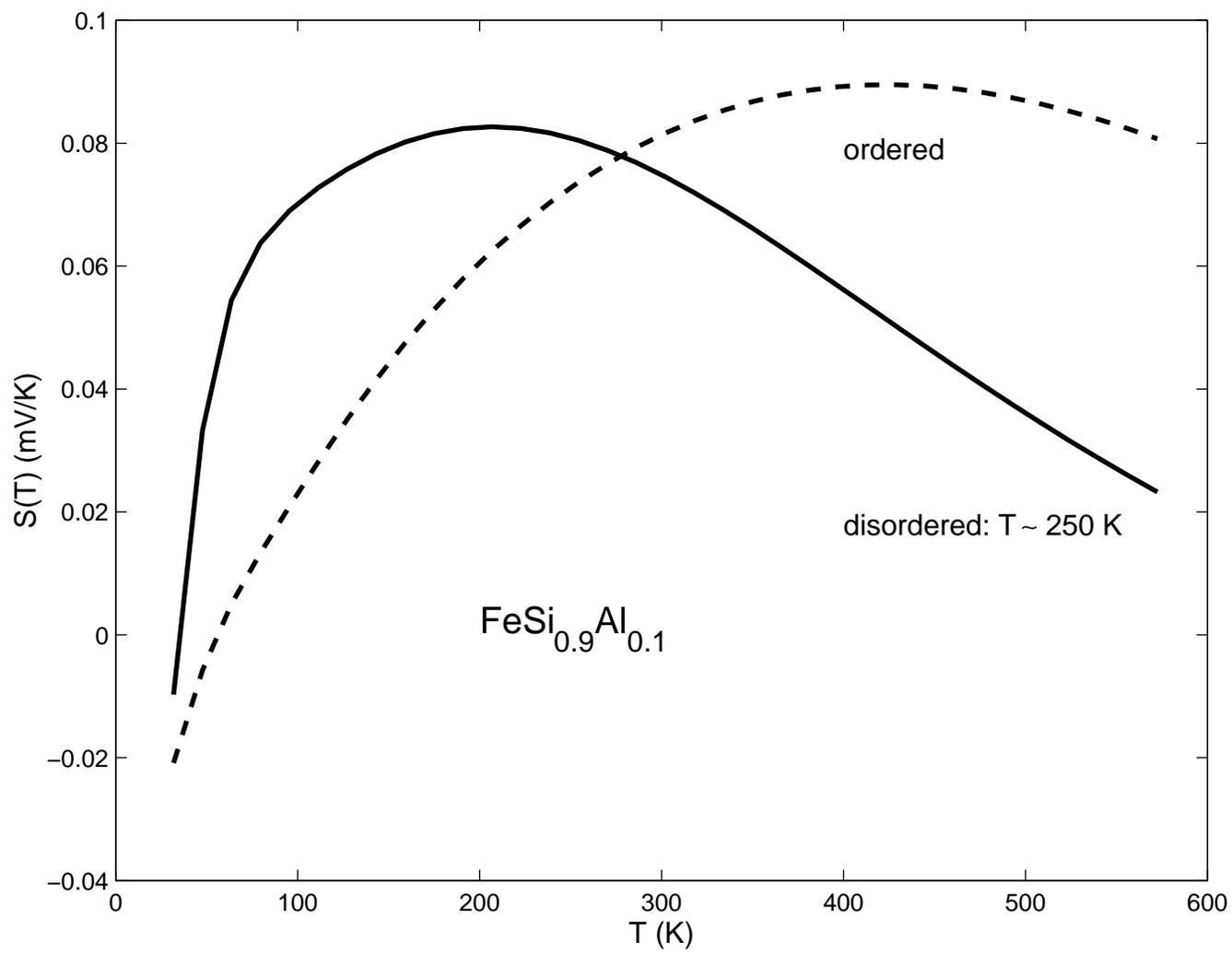
\end{document}